\documentclass[12pt]{article}

\newcommand{\be}{\begin{equation}}
\newcommand{\ee}{\end{equation}}
\newcommand{\beq}{\begin{eqnarray}}
\newcommand{\eeq}{\end{eqnarray}}
\newcommand{\bc}{\begin{center}}
\newcommand{\ec}{\end{center}}

\begin{document}

\title{\bf The theory of relativity \\and\\ the Pythagorean theorem\footnotemark 
}
\author{\bf L. B. Okun}
\date{}
\renewcommand\thefootnote{\fnsymbol{footnote}}
\maketitle

\abstract
It is shown that the most important effects of special and general theory of relativity can be understood in a simple and straightforward way.
The system of units in which the speed of light $c$ is the unit of velocity allows to cast all formulas in a very simple form.The Pythagorean theorem graphically relates energy, momentum and mass. The paper is addressed to those who teach and popularize the theory of relativity

\section{Introduction}

The report "Energy and mass in the works of Einstein, Landau and Feynman" that I was preparing for the Session of the Division of Physical Sciences of the Russian Academy of Sciences (DPS RAS) on the occasion of the 100th anniversary of Lev Davidovich Landau's birth was to consist of two parts, one on history and the other on physics. The history part was absorbed into the article ``Einstein's formula: $E_0=mc^2$. `Isn't the Lord laughing?''' that appeared in the May issue of Uspekhi Fizicheskikh Nauk [Physics-Uspekhi] journal [1]. The physics part is published in the present article. It is devoted to various, so to speak, technical aspects of the theory, such as the dimensional analysis and fundamental constants $c$ and $\hbar$; the kinematics of a single particle in the entire velocity range from 0 to $c$; systems of two or more free particles; and the interactions between particles: electromagnetic, gravitational, etc. The text uses the slides of the talk at the session of the Section of Nuclear Physics of the DPS RAS in November 2007 at the Institute for Theoretical and Experimental Physics (ITEP). My goal was to present the main formulas of the theory of relativity in the simplest possible way, using mostly the Pythagorean theorem.

\footnotetext[1]{This is a slightly corrected version of the paper published in {\it Physics - Uspekhi {\bf 51} 622 (2008).} In particular, the short
subsections of the text are numbered as had been suggested by one of the readers.}

\renewcommand{\thefootnote}{\arabic{footnote}}

\section{Relativity}

{\bf 2.1. The advanced standpoint.}

The history of the concept of mass in physics runs to many centuries and is very interesting, but I leave it aside here. Instead, this will be an attempt to look at mass from an advanced standpoint. I borrowed the words from the famous title of Felix Klein's {\it Elementary Mathematics from an Advanced Standpoint}
(traditionally translated into Russian incorrectly as {\it Elementary Mathematics from the Standpoint of Higher Mathematics.} See V.G. Boltyanskii's foreword to the 4th Russian edition). 
The advanced modern standpoint based on principles of symmetry in general and on the theory of relativity in particular makes it possible to avoid inevitable terminological confusion and paradoxes.

{\bf 2.2. The principle of relativity.}
Ever since the time of Galileo and Newton, the concept of relativity has been connected with the impossibility of detecting, by means of any experiment, a translational (uniform and rectilinear) motion of a closed space (for instance, inside a ship) while remaining within this space.
At the turn of XIX and XX
 centuries Poincar\'{e} gave to this idea the name `the principle of relativity'  . In 1905 Einstein generalized this principle to the case of the existence of the limiting velocity of propagation of signals. (The finite velocity of propagation of light has been discovered by
R$\ddot{\rm o}$mer already in 1676). Planck called the theory constructed in this way `Einstein's theory of relativity'.

{\bf 2.3. Mechanics and optics.} Newton tried to construct a unified theory uniting the theory of motion of massive objects (mechanics) and the theory of propagation of light (optics). In fact, it became possible to create the unified theory of particles of massive matter and of light only in the XX century. It was established on the road to the vantage ground of truth (I am using here the ironical wording of Francis Bacon) that light is also a sort of matter, just like the massive stuff, but that its particles are massless. This interpretation of particles of light --- photons --- continues to face resistance from many students of physics, and even more from physics teachers.

\section{Dimensions}

{\bf 3.1. Units in which c = 1.} The maximum possible velocity is known as the speed of light and is denoted by $c$. When dealing with formulas of the theory of relativity it is convenient to use a system of units in which $c$ is chosen as a unit of velocity. Since $c/c=1$, using this system means that we set $c = 1$ in all formulas, thus simplifying them greatly. If time is measured in seconds, then distance in this system of units should be measured in light seconds: one light second equals $3 \cdot 10^{10}$  cm.

{\bf 3.2. Poincar\'{e} and c.} One of the creators of the theory of relativity, Henri Poincare, when discussing in 1904 the fact that $c$ is found in every equation of electrodynamics, compared the situation with the geocentric theory of Ptolemy's epicycles in which every relation between motions of celestial bodies included the terrestrial year. Poincare expressed his hope that the future Copernicus would rid electrodynamics of $c$ [3]. However, Einstein showed already in 1905 that $c$ was to play the key role as the limit for the velocity of signal propagation.

{\bf 3.3. Two systems of units: SI and c = 1.} The unit of velocity in the International System of Units SI, 1 m/s, is forced on us by convenience arguments and by standardization of manufacturing and commerce but not by the laws of Nature. In contrast to this, $c$ as a unit of velocity is imposed by Nature itself when we wish to consider fundamental processes of Nature.

{\bf 3.4. Dimensional factors.} Consider some physical quantity $a$. Let us denote by [$a$] the dimension of the quantity $a$. The dimension of $a$ definitely changes if it is multiplied by any power of the universal constant $c$ but its physical meaning remains unaffected. In what follows I explain why this is so.

{\bf 3.5. Velocity, momentum, energy, mass.} The dimensions of momentum, mass, and velocity of a particle are usually related by the formula
$[\bf p] = [m][\bf v]$ while the dimensions of energy, mass, and velocity are related by the formula
$[E\,] \!=\! [m][{\bf v}^{\,2}]$.
Let us introduce dimensionless velocity ${\bf v}/c$ and from now on denote this ratio as $\bf v$. Likewise, referring to momentum $\bf p$ we actually mean the ratio ${\bf p}/c$. When speaking of energy, we actually mean the ratio $e = E/c^{\,2}$. Obviously, the dimensions of $\bf p$, $e$, and $m$ become identical and therefore, these quantities can be measured in the same units, for example, in grams or electron-volts, as is customary in elementary particle physics.

{\bf 3.6. On the letter e denoting energy.} Choosing $e$ as the notation for energy may invite the reader's ire since this symbol traditionally stands for electron and electric charge. However, this choice cannot cause confusion and, importantly, it will lead to a compact form of formulas for a single particle, always reminding us that these formulas were written using the system of units in which $c=1$. On the other hand, it will be clear a little later that the letter $E$ is a convenient notation for the energy of two or more particles.
I happened to see Einstein's formula with a lower-case $e$ on a billboard on Rublevskoye highway in Moscow. I wonder, why should this $e$ irritate physicists?

{\bf 3.7. On the difference between energy and frequency.} Two paragraphs ago I insisted that $e=E/c^{\,2}$ is energy even though its dimension is that of mass. In that case it is logical to ask why $\omega = E/\hbar$  is not energy but frequency? Indeed, the quantum of action $\hbar$, like the speed of light $c$, is a universal constant. The answer to this question can be found by considering how $e$ and $\omega$ are measured. $E$ and $e$ are measured by the same procedure, say, using a calorimeter, while frequency is measured in a drastically different manner, say, using clocks. Therefore, the equality $\omega =E/\hbar$  informs us of the link between two different types of measurement, while the equality $e=E/c^{\,2}$ carries no such information. Arguments similar to those concerning frequency hold equally well for wavelength. I have to emphasize that these metrological distinctions are mostly of a historical nature since in our day atomic clocks operate on the difference between atomic energy levels.

\section{Single particle}

{\bf 4.1. Relative and absolute quantities.} The kinetic energy of any body is a relative quantity: it depends on the reference frame in which it is measured. The same is true for the momentum of a body and its velocity. In contrast to them, the mass of a body is an absolute quantity: it characterizes the body as such, irrespective of the observer. The rest energy of a body (see below) is also an absolute quantity since the frame of reference is fixed in it once and for all --- `nailed to it'.

{\bf 4.2. Invariant mass.} The mass of a body is defined in the theory of relativity by the formula
\be m^2 = e^2 - p^2.\ee
Here and in what follows $p = |{\bf p}|$. Likewise, $v = |{\bf v}|$.
Note that energy and momentum of a given body are not bounded from above while the mass of the body is fixed. Formula (1) is the simplest relation between energy, momentum, and mass that one could write `off the top of one's head'. (The relation between $e$, $\bf p$, and $m$ cannot be linear since $\bf p$ is a vector while $e$ and $m$ are scalars in three-dimensional space.) We shall see now that formula (1) has another, much more profound theoretical foundation.

{\bf 4.3. The 4-momentum.} Minkowski was the first to point out that the theory of relativity gains the simplest form if considered in four-dimensional spacetime [4]. Energy and momentum in the theory of relativity form a four-dimensional energy-momentum vector $p_i (i=0,a)$, where $p_0= e$, $p_a= {\bf p}$, and $a=1,2,3$.
Mass is the Lorentz scalar that characterizes the length of the 4-vector $p_i$:  $m^{\,2}={p_i}^2=e^{\,2}- {\bf p^2}$; four-dimensional space is pseudo-Euclidean, which explains the minus sign in the formula for length squared. (The reader will recall that ${\bf p}^2=p^2$.) Another way to clarify why the sign is negative is by introducing the imaginary momentum ${\rm i}{\bf p}$. Then $m^{\,2}= e^{\,2} + (i{\bf p})^2$ and we are dealing with the Pythagorean theorem for such a pseudo-Euclidean right triangle in which the hypotenuse $m$ is shorter than the cathetus $e$.

{\bf 4.4. Relation between momentum and velocity.} The momentum of a body is related to its velocity $\bf v$ by the formula
\be  {\bf p}= e{\bf v}.\ee
This formula satisfies in the simplest manner the requirement that the momentum 3-vector be proportional to the velocity 3-vector and that the dimensional proportionality coefficient not vanish for the massless photon.
Conservation of the thus defined momentum in the theory of relativity is implied by the uniformity of 3-space while conservation of energy is implied by the uniformity of time (Noether's theorem).

{\bf 4.5. The Pythagorean theorem.} Formula (1) is shown in Fig. 1 by an ordinary right triangle in which $m$ and $p$ are catheti and $e$ is the hypotenuse.

{\bf 4.6. Transition from  ${\bf m}\neq {\bf 0}$ to ${\bf m}={\bf 0}$.} Formula (1) is obviously valid at $m = 0$ while formula (2) holds for $v=1$. This implies that there is a smooth transition from massless particles to massive, when the energy of the latter particles greatly exceeds their mass.

{\bf 4.7. Physics from ${\bf p}={\bf 0}$ to ${\bf p}={\bf e}$.} Let us consider formulas (1) and (2) first at zero momentum, then in the limit of very low momenta (when $p\ll m$), and then in the limit of very high momenta when $p \sim e \gg m$, and finally in the case of massless photons.
We will call the case of very small momenta and velocities the Newtonian case, and that of very high momenta and velocities close to the speed of light, the ultrarelativistic case. We will start with zero momentum.

%
%

\begin{center}
\unitlength 2mm 
\linethickness{0.8pt}
\ifx\plotpoint\undefined\newsavebox{\plotpoint}\fi 
\begin{picture}(44.25,48.5)(0,0)
\put(9.25,48.5){\line(0,-1){29}}
\put(9.25,19.5){\line(1,0){35}}
\multiput(44.25,19.5)(-.0409883721,.0337209302){860}{\line(-1,0){.0409883721}}
\put(7.5,33.25){\makebox(0,0)[cc]{$p$}}
\put(25.25,36.75){\makebox(0,0)[cc]{$e$}}
\put(22.5,18){\makebox(0,0)[cc]{$m$}}
\put(22.25,10.75){\makebox(0,0)[cc]{Fig. 1}}
\end{picture}
\end{center}

\section{Rest energy}

{\bf 5.1. Zero momentum.} If momentum is zero, then in the case of a massive particle the velocity is also zero and energy $e$ is by definition equal to the rest energy $e_0$. (The subscript 0 reminds us that here we are dealing not with the energy of a given body in general but with its energy precisely in the case when its momentum is zero!) Hence equation (1) implies
\be e_0=m. \ee
If, however, the particle is massless, then equation (1) at $p=0$ implies that  $e=e_0=0$ (see 7.6).

{\bf 5.2. Horizontal `biangle'.} If $m \neq 0$ and $p=0$, then the triangle shown in Fig.1 `collapses' to a horizontal `biangle' (Fig. 2).

{\bf 5.3 Einstein's great discovery.} In units in which $c\ne 1$, equation (3) has the form
\be E_0= mc^{\,2} \,.\ee
The realization that ordinary matter at rest stores an enormous amount of energy in its mass was Einstein's great discovery.

{\bf 5.4. The `famous formula'.} Equation (4) is very often written (especially in popular physics literature) in the form of `Einstein's famous equation' that drops the subscript 0:
\be E=mc^{\,2}\,.	\ee
This simplification, to which Einstein himself sometimes resorted, might seem innocuous at first glance, but it results in unacceptable confusion in understanding the foundations of physics. In particular, it generates a totally false idea that `according to the theory of relativity' the mass of a body is equivalent to its total energy and, as an inevitable result, depends on its velocity. (`Wished to make it simpler, got it as always'.\footnote{A paraphrase of former Russian Prime Minister Chernomyrdin's `statement of the day': ``Wished to make it better, got as always.'' (Note added by the Author in translation.)})

%
%

\begin{center}
%
\unitlength 2mm 
\linethickness{0.8pt}
\ifx\plotpoint\undefined\newsavebox{\plotpoint}\fi 
\begin{picture}(47,23.5)(0,0)
\put(10.5,18.25){\line(1,0){30.5}}
\put(26.25,20.5){\makebox(0,0)[cc]{$e_0$}}
\put(26.25,16.25){\makebox(0,0)[cc]{$m$}}
\put(26.25,9.5){\makebox(0,0)[cc]{Fig. 2}}
\end{picture}
\end{center}

{\bf 5.5. No experiment can disprove the `famous formula'.} Very clever people thought up this formula in such a way that it never contradicts experiments. However, it contradicts the essence of the theory of relativity. In this respect, the situation with the `famous formula' is unique --- I do not know another case that could be compared with this one.

{\bf 5.6. This is not a matter of taste but of understanding.} You hear time and again that the introduction of momentum-dependent mass is `a matter of taste'. Of course, one can write the letter $m$ instead of $E/c^{\,2}$ and even call it `mass', although it is no more sensible than writing $p$ instead of $E/c$ and calling it `momentum'. Alas, this `dress changing' introduces unnecessary and bizarre notions --- relativistic mass and rest mass $m_0Т$--- and creates an obstacle to understanding the theory of relativity. A well-known Russian proverb comes to mind: ``Call me a pot if you wish but don't push me into the oven.'' Unfortunately, people who call $E/c^{\,2}$ `mass' do place this `pot' into the `oven' of physics teaching.

{\bf 5.7. Longitudinal and transverse masses.} In addition to relativistic mass, concepts of intense use at the beginning of the XX century were the transverse and longitudinal masses: $m_{\rm t}$ and $m_{\rm l}$. This longitudinal masses increased as $(e^{\,3}/m^{\,3})\,m$ and `explained' --- in terms of Newton's formula $F=ma$ --- why a massive body cannot be accelerated to the speed of light. Then it was forgotten and such popularizers of the theory of relativity as Stephen Hawking started to persuade their readers that even much gentler growth of mass with velocity $((e/m)\,m)$ could explain why the velocity of a massive body cannot reach $c$. I single out Hawking only because, printed on the dust jacket of the Russian edition of his latest popular science book [5], which advertises the formula $E=mc^{\,2}$, we see this text: ``Translated into 40 languages. More than 10 million copies sold worldwide.''

{\bf 5.8. False intuition.} After my talk at the ITEP A N Skrinsky told me that the notion of relativistic mass hampered a well-known physicist's understanding that a relativistic electron colliding with an electron at rest can transfer all its energy to the latter. Indeed, how could a heavy baseball bat transfer all its energy to the lightest ping-pong ball? In physics, as in daily life, people very often rely on intuition. This is why it is so important, when studying the theory of relativity, to work out the relativistic intuition and mistrust nonrelativistic intuition. (In order to `feel' how an electron at rest can receive the entire energy of a moving electron it is sufficient to use their center-of-inertia frame to consider scattering by 180 degrees, and then return back to the laboratory frame.)

\section{Newtonian mechanics}

{\bf 6.1. Momentum in Newtonian mechanics.} Newtonian mechanics describes with high accuracy the motion of macroscopic bodies in a terrestrial environment and of massive celestial bodies because their velocities are much smaller than the speed of light. For instance, the velocity of a bullet is of the order of 1 km/s, which corresponds to $ v = 1/300 000$ and $v^{\,2} = 10^{-11}$. In this situation equation (2) reduces to
\be {\bf p} = m{\bf v} \,.\ee
Equation (1) is schematically shown in the Newtonian limit in Fig. 3.
The side of the triangle representing $p$ in Fig. 3 is far too long. Scaled correctly, it should be a few microns.

%
%
%
%

\begin{center}
\unitlength 2mm 
\linethickness{0.8pt}
\ifx\plotpoint\undefined\newsavebox{\plotpoint}\fi 
\begin{picture}(48.25,26)(0,0)
\put(9.75,26){\line(0,-1){8.25}}
\put(9.75,17.75){\line(1,0){38.5}}
\multiput(48.25,17.75)(-.157142857,.033673469){245}{\line(-1,0){.157142857}}
\put(8.,22.25){\makebox(0,0)[cc]{$p$}}
\put(25.5,24.5){\makebox(0,0)[cc]{$e$}}
\put(23.25,16.25){\makebox(0,0)[cc]{$m$}}
\put(22.5,7.75){\makebox(0,0)[cc]{Fig. 3}}
\end{picture}
\end{center}

{\bf 6.2. Kinetic energy ${\bf e}_{\bf k}$.} It is reasonable to rewrite formula (1) for low velocities so as to isolate the contribution of the short cathetus:
\be e^{\,2} - m^{\,2} = p^{\,2}\,	\ee
and then to present it in the form
\be (e-m)(e+m) = p^{\,2} \,.\ee
This allows us to obtain a nonrelativistic expression for kinetic energy without resorting to the conventional series expansion of the square root. We take into account that the total energy $e$ is the sum of rest energy $e_0$ and kinetic energy $e_{\rm k}$ and therefore $e= m+ e_{\rm k}$.

{\bf 6.3. Energy in Newtonian mechanics.} In the Newtonian limit we have $e_{\rm k} \ll m$ (e.g. for a bullet  $e_{\rm k}/ m = 10^{-11}$). Energy can therefore be replaced with high accuracy by mass $m$ in formula (2) for momentum and in the factor $(e+m)$  in equation (8). This last equation immediately implies an expression for kinetic energy $e_{\rm k}$ in Newtonian mechanics:
\be e_{\rm k} = {p^{\,2}\over 2m} = {mv^{\,2}\over 2} \,. \ee

{\bf 6.4. Potential energy.} In addition to velocity-dependent kinetic energy, an important role in nonrelativistic mechanics is played by potential energy, which depends only on the position (coordinate) of the body. The sum of kinetic and potential energy is conserved at any instance of time. 
The potential energy of a body placed in an external field of force is defined to within an arbitrary additive constant because the force acting on the body equals the gradient of potential energy. In a similar manner, the potential energy of interaction of several bodies depends only on their positions at the moment of interaction. However, in the theory of relativity any interaction propagates at a finite velocity. Hence, potential energy is an essentially nonrelativistic concept.

{\bf 6.5. Newton and modern physics.} Newton's flash of genius marked the birth of modern science. The post-Newtonian progress of science is fantastic. Today's understanding of the structure of matter is radically different from Newton's. Nevertheless, even in the XXI century many physics textbooks continue to use Newton's equations at energies $e_{\rm k} \gg e_0$, which exceed the limits of applicability of Newton's mechanics ($e_{\rm k} \ll e_0$) by many orders of magnitude.
If some professors prefer to insist on keeping up with this tradition of velocity-dependent mass, they ought to at least familiarize their students with the fundamental concepts of mass and rest energy, and with the true Einstein equation $E_0=mc^{\,2}$.

\section{Ultrarelativism}

{\bf 7.1. High energy physics.} Let us now consider in some detail the limiting case in which $e/m \gg 1$. The ratio of energy and mass characteristic for high energy physics is precisely this. For example, this ratio for electrons in the LEP (Large Electron-Positron) Collider at CERN was $e/m = 10^5$, since $m=0.5$ MeV and $e=50$ GeV. For protons in the LHC (Large Hadron Collider), which is located in the same tunnel where the LEP was in previous years, we find $e/m \sim 10^4$. (Here, $m \sim 938$ MeV, $e \sim 7$ TeV.)

{\bf 7.2. A vertical triangle.} The triangle for protons in the LHC is drawn highly schematically in Fig. 4. Its base is 
in fact
shorter than its hypotenuse by four orders of magnitude.

{\bf 7.3. The neutrino.} Neutrinos are even more ultrarelativistic particles: their masses are a fraction of one electron-volt and their energies reach several MeV for neutrinos emerging from the Sun and nuclear reactors, and several GeV for neutrinos generated in particle decays in cosmic rays and in accelerators. The base of the triangle shown schematically in Fig. 4 is much shorter at these energies than its vertical cathetus and its hypotenuse.


\begin{center}
%
\unitlength 2mm 
\linethickness{0.8pt}
\ifx\plotpoint\undefined\newsavebox{\plotpoint}\fi 
\begin{picture}(31.25,57)(0,0)
\put(23,57){\line(0,-1){34.25}}
\put(23,22.75){\line(1,0){7.25}}
\multiput(30,22.75)(-.03365385,.16466346){208}{\line(0,1){.16466346}}
\put(21.25,37.5){\makebox(0,0)[cc]{$p$}}
\put(28.25,38.75){\makebox(0,0)[cc]{$e$}}
\put(26.5,21.5){\makebox(0,0)[cc]{$m$}}
\put(25.75,15.25){\makebox(0,0)[cc]{Fig. 4}}
\end{picture}
\end{center}

{\bf 7.4. Neutrino oscillations and ${\bf m}^{\bf 2}/{\bf 2}{\bf e}$.} Equation $(e\!-\!p)(e\!+\!p)\!=m^{\,2}$ immediately implies that $e-p \simeq m^{\,2}/2e$. The differences between the masses of three neutrinos $\nu_1, \nu_2, \nu_3$ possessing definite masses in a vacuum result in oscillations between neutrinos having no well-defined masses but possessing certain flavors: ${\nu }_{\rm e}, {\nu }_{\mu }, {\nu }_{\tau }$. (This phenomenon is similar to well-known beats that occur when several frequencies interfere.) The neutrino oscillation data give
$$	\Delta m^{\,2}_{21} = (0.77\pm 0.04)\times 10^{-4}\ {\rm eV}^2\,$$

 $$|\Delta m^{\,2}_{32}| = (24\pm 3)\times 10^{-4}\ {\rm eV}^2\,.$$

{\bf 7.5. The photon.} The photon mass is so small that no experiment has been able to detect it. Hence, it is usually assumed that the photon mass equals zero. This means that for a photon $e=p$, where $p={|\bf p|}$, and the triangle shown in Fig. 4 collapses to a vertical biangle (Fig. 5).

%
%

\begin{center}
\unitlength 2mm 
\linethickness{0.8pt}
\ifx\plotpoint\undefined\newsavebox{\plotpoint}\fi 
\begin{picture}(29.43,55)(0,0)
\put(20.5,55){\line(0,-1){28}}
\put(19,38.75){\makebox(0,0)[cc]{$p$}}
\put(22.,38.8){\makebox(0,0)[cc]{$e$}}
\put(20.6,21.61){\makebox(0,0)[cc]{Fig. 5}}
\end{picture}
\end{center}

{\bf 7.6 The photon and rest energy?} It is logical to conclude the discussion of single-particle mechanics by returning to the question: is the concept of rest energy $e_0$ applicable to massless photon?
It may seem at first glance that it is not, since a photon propagates at the speed $c$, however small its energy is, so that `a rest for it is but a dream' \footnote{This is a paraphrase of the famous line from Alexander Block. (Note added in translation).}. This being so, how can we use the equality $e_0=0$ if the photon is never at rest? 
We can because our
$e_0$ is 
defined as 
the energy corresponding to zero momentum,
not velocity. Obviously this energy is zero for the photon with $p=0$: this is implied by equation (1). 
If a particle has $m = 0, p=0, e=0$ and biangle of Fig. 5 collapses to a point, we can say that it `passed away to the state of eternal rest'. Looking at the limiting transition to zero mass, we can show that the reference frame in which a photon is `eternally at rest' has to be rigidly connected to another `eternally resting' photon. Consequently, the value $e_0=0$ at $m=0$ is in perfect agreement with the limiting transition.

\section{Two free particles}

{\bf 8.1. Collision of two particles. Colliders.} If two particles collide at relativistic energies, a comparison of the reference frame in which one of them is at rest with a reference frame in which their common center of inertia is at rest demonstrates the advantages of the latter. We already saw this in the case commented on by A N Skrinsky. If the momenta of the colliding particles are equal and oppositely directed, as for example in the LHC or LEP collider, then practically the entire energy of the colliding particles may be spent on the creation of new particles.

{\bf 8.2. Mass of a system of particles.} The total energy $E$ and the total momentum $\bf P$ of an isolated system of particles are conserved. Energy and momentum being additive, for two free particles we have
\be	E = e_1 + e_2 \, \ee

\be {\bf P} = {\bf p}_1 + {\bf p}_2. \ee
We now define the quantity $M$ by the formula
\be M^{\,2} = E^{\,2} - {\bf P}^{\,2}.\ee

{\bf 8.3. Masses are additive at v = 0.} Equation (12) is invariant under Lorentz transformations, as is equation (1). Therefore, it is logical to refer to $M$ as the mass of a system of two particles. In the static limit, when $p_1$ and $p_2$ equal zero, equation (12) implies that
\be M = e_{01} + e_{02} = m_1 + m_2. \ee
In the Newtonian limit, $M$ equals the sum of the masses of the two particles with an accuracy of $(v/c)^2$, i.e. the masses are practically additive.

{\bf 8.4. Masses are not additive at  v $\ne$ 0.}  However, $M$ and the masses $m_1$ and $m_2$ are practically unrelated at high velocities. For instance, $M$ exceeds the electron mass in the LEP collider or the proton mass in the LHC by four orders of magnitude (see section 7). The value of $M$ is crucially dependent on the relative directions of the momenta of two particles, since the sum of two vectors is a function of the angle between them. Thus, we have for two photons moving in the same direction
\be P = |{\bf P}| = |{\bf p}_1 + {\bf p}_2| = p_1 + p_2.\ee

{\bf 8.5. Collinear photons.} For photons $p_1=e_1$, and $p_2=e_2$. Therefore, for two photons moving in the same direction we can write
\be P = p_1 + p_2 = e_1 + e_2 = E.\ee
Equation (12) then implies that in this case the mass of a pair of photons $M=0$. And this means that the mass of a `needle' light beam is zero.

{\bf 8.6. What if photons fly away from each other?} However, if photons fly away in opposite directions with equal energies, then ${\bf p}_1 = -{\bf p}_2$ and ${\bf P} = 0$. In that case, the rest energy of two photons simply equals the sum of their energies and the mass of this system is
\be M = E_0 = 2e.\ee

{\bf 8.7. Shock.} Of course, the statement that a pair of two massless or very light particles might have an enormous mass may shock the unprepared reader. Is there any sense in speaking of the rest energy of two photons if `rest is but a dream' to either of them? What is at rest in this case?

{\bf 8.8. The answer is obvious.} The entity at rest is the geometric point -- the center of inertia of the two photons. While the rest energy for one particle is the energy hidden in its mass, for two photons it is simply the sum of their energies (kinetic energies!) in the reference frame in which their momenta are equal in magnitude and opposite in direction. There is no hidden energy in this case!

{\bf 8.9. What does it mean to be conserved?} When saying that energy is conserved, we mean that the sum of the energies of particles entering a reaction equals the sum of the energies of particles created as a result of this reaction. The statement on the conservation of momentum has a similar meaning. However, since momentum is a vector quantity, now we are dealing with a vector sum of momenta. (In the case of momenta we speak about three independent conservation laws: conserved are the sums of projections of momenta on three mutually orthogonal directions.)
The conserved quantities are thus $E= \sum e_i$ and ${\bf P} =\sum {\bf p}_i $. As for the energies of individual particles $e_i$ their momenta ${\bf p}_i$ in the laboratory reference frame, they are conserved only in elastic forward scattering.
Here, it is important to stress the difference between the concepts of additivity and conservation. The former concept refers to the state of a system of free particles, the latter refers to the process of interaction of the particles.

{\bf 8.10. Is mass conserved?} With $E$ and $\bf P$ conserved, the mass $M$ of a system (a set) of particles, defined by the formula $M^{\,2}=E^{\,2}-{\bf P}^{\,2}$, must be conserved as well. In contrast to energy and momentum, however, mass is not additive: $M \neq \sum m_i$. Some authors talk about the non-additivity of mass as if it were identical to its non-conservation (e.g. we find this statement in  \S 9 of {\it Field Theory} by Landau and Lifshitz [6].) In fact, as I emphasized above, in general neither masses nor energies or momenta are conserved for individual particles participating in a reaction; not even the particles themselves are. Hence, it is incorrect to speak of mass nonconservation as something in contrast to conservation of energy and momentum.

{\bf 8.11. Einstein's thought experiment.} Of course, the concept of the mass of two photons flying away from each other looks rather strange. However, it was by using this very idea that Einstein came to discover the rest energy of a massive body in 1905. He noticed that having emitted `two amounts of light' in opposite directions, the body at rest continues to stay at rest but that its mass in this thought experiment diminishes. In the laboratory reference frame both the body and the center of inertia of the two photons are at rest. Consequently, the mass of the initial body equals the sum of two masses: that of the resulting body and that of the system of two photons.

{\bf 8.12. Positronium annihilation.} {\em Nihil} in Latin means {\em nothing}. A positronium is an `atom' consisting of a positron and an electron. The reaction in which a positronium converts to two photons ${\rm e^+ e^-} \to \gamma \gamma$  was given the name annihilation, perhaps because at that time photons were not considered particles of matter. Annihilation conserves $M$ because $E$ and $\bf P$ are conserved. In the initial state $M$ equals the sum of masses of the electron and the positron [minus the binding energy, which is small and in this context irrelevant (see below)]. In the final state $M$ equals the sum of energies of two photons in the positronium's rest frame. The rest energy of the electron and the positron thus transforms completely into the energy (kinetic) of the photons, but the masses of the initial and final states are identical in this process, exactly as follows from the conservation of total energy and total momentum.

{\bf 8.13. Meson decays.} Likewise, when a K meson decays into two or three $\pi$ mesons, the kaon's rest energy transforms into the sum of total energies of the pions, each of which has the form $e = e_{\rm k} + m$. However, the mass of a system of two or three pions produced in the decay of a kaon equals the kaon mass.

{\bf 8.14. What do we call `matter'?} In any decay the rest energy transforms into the energy of motion, while the total energy of an isolated system remains conserved. The mass of the system is conserved but the masses of its individual particles are not. Massive particles decay into less massive particles, or sometimes into massless ones. In elementary particle physics we call  `particles of matter'  not only massive particles such as protons and electrons, but also very light neutrinos and massless photons, and even gravitons (see below). Today's quantum field theory treats all of them on an equal basis.

{\bf 8.15. Energy without particles?} Matter does not disappear in decay and annihilation reactions leaving behind only energy like the Cheshire cat would leave behind only its smile. In all these processes the carriers of energy are particles of matter. Energy without matter (`pure energy') has never been observed in any process studied so far.
True, this might be not so for so-called dark energy, which was discovered in the last years of the XX century. Dark energy manifests itself in the accelerating expansion of the Universe. (The evidence for this accelerating expansion is found in recession velocities of remote supernovas.) Three-fourths of the entire energy in the Universe is dark energy and its carrier appears to be the vacuum. The remaining quarter is carried by ordinary matter (5\%) and dark matter (20\%). Dark energy does not affect processes with ordinary matter observed in laboratories. In a laboratory experiment energy is always carried by particles.

\section{Non-free particles}

{\bf 9.1. Bodies and particles.} All physical bodies consist of elementary particles. Such elementary particles as the proton and the neutron are themselves made up of `more elementary particles' --- quarks and gluons. Such particles as the electron and the neutrino appear at our current level of understanding as truly elementary particles. The feature common for the proton and the electron is that the masses of all protons in the world are strictly identical, as are the masses of all electrons. In contrast to this, the masses of all macroscopic bodies of the same type, say, of all 10-cent coins, are only approximately equal.
Practically the difference between two coins arises because the process of minting coins is far from being ideal. What is more important here is that the mass of a coin is
not well defined because different energy levels of a coin are practically degenerate, while the mass of the nearest excited state of a proton exceeds the proton mass by several hundred MeV. Therefore Nature mints ideally identical protons. 

{\bf 9.2. Mass of a gas.} In all the cases discussed above, particles moved away freely when the mass of the system of particles was greater than the sum of their masses. Let us turn now to a situation in which they are not free to move away. This situation is found, for example, in the frequently discussed thought experiment with a gas of molecules or photons in a closed vessel at rest. The total momentum of this gas is zero because the gas is isotropic: ${\bf P} =\sum {\bf p}_i =0$. Hence, the total mass $M$ of this gas equals its total energy $E$ (and in this case it is identical to $E_0$) and hence to the sum of energies of individual particles: $M = E= \sum e_i$.

{\bf 9.3. Mass of a heated gas.} When gas in a nonmoving vessel is heated, its total momentum remains unchanged and equal to zero while the total energy increases because the kinetic energy of every particle increases. As a result, the mass of the gas as a whole increases, while the mass of each individual particle remains unchanged. (Sometimes a wrong statement may be encountered in the literature that the masses of particles (or photons) increase as their kinetic energies are increased.)

{\bf 9.4. Mass of a hot iron.} In the same manner, the mass of an iron must increase as it heats up, even though the masses of the vibrating atoms remain the same. However, the set of formulas (10)--(12) written for a system of free particles cannot be applied to the iron since the particles (atoms in this case) are not free but are tied into the crystal lattice of the metal. Obviously, an increase in the iron mass is too small to be measurable.

\section{Atoms and atomic nuclei}

{\bf 10.1. On formulas (10)--(12).} Why are formulas (10)--(12) unsuitДable for dealing with such non-free particles as electrons in atoms and nucleons in atomic nuclei? First and foremost, on account of the uncertainty relation these particles do not possess precisely defined momenta. The smaller the volume to which they are confined, the greater is the uncertainty of their momenta.

{\bf 10.2. Uncertainty relation.} The laws of quantum mechanics, and the uncertainty relation as one among them, are very important both for atoms and for nuclei. As we know, the product of the momentum uncertainty $\Delta p$ and the coordinate uncertainty $\Delta x$ must be not smaller than the quantum of action $\hbar$. Hence, particles within atoms have no definite momenta and only possess a certain total momentum.

{\bf 10.3. Energy of the field.} Another reason why formulas (10)--(12) are not valid inside atoms is the fact that the space between individual particles in an atom is essentially not empty but filled with a material medium, i.e. physical fields. The space inside the atom is filled with an electromagnetic field and the space inside a nucleus, by a much denser and stronger field, often described as the meson field.

{\bf 10.4. Real and virtual particles.} In classical theory particles and fields are concepts that cannot be reduced to one another. In quantum field theory we use the language of Feynman diagrams, which reduce the concept of a field to that of a virtual particle for which  $e^{\,2} - {\bf p}^{\,2} \ne  m^{\,2}$. We say about such particles that they are off mass shell. (Particles that are called on mass shell are real particles and for them $e^{\,2} -{\bf p}^{\,2} = m^{\,2}.)$ Also, the 4-momentum $p_i = (e,{\bf p})$ is conserved at each vertex of the diagram.

{\bf 10.5. Binding energy.} As a result of the presence of the field, we need to take into account in formula (10), $E=e_1+e_2$, the field energy of two closely interacting particles, say, in the deuteron, the nucleus of heavy hydrogen. Consequently, $M<m_1+m_2$. The quantity $\varepsilon = m_1\,+\,m_2-M$ is known as the binding energy. The mass of the deuteron is less than the mass of the proton plus that of the neutron of which deuteron consists. The binding energy of nucleons in deuteron is 2.2 MeV. To break deuteron into nucleons we need to spend an amount of energy equal to or greater than the binding energy. The atomic nuclei of all other elements of the periodic Mendeleyev table also owe their existence to the binding energy of their nucleons in the nucleus.

{\bf 10.6. Fusion and fission of nuclei.} We know that the binding energy 
per nucleon 
rises to a maximum at the beginning of the periodic Mendeleyev table for the helium nucleus and in the middle of the Table for the iron nucleus. This is why huge amounts of kinetic energy are released when helium is formed from hydrogen in fusion reactions in the Sun and in hydrogen bombs. In nuclear reactors and atomic bombs, kinetic energy is released by fission reactions when heavy nuclei of uranium and plutonium break into lighter nuclei from the middle of the periodic Mendeleyev Table.

{\bf 10.7. Chemical reactions.} Substantially lower energy, on the order of electron-volts, is released in chemical reactions. It is caused by differences in binding energies in various chemical compounds. However, the source of kinetic energy in both chemical and nuclear reactions is the difference between the masses of initial and final particles (molecules or nuclei) that take part in these reactions.
Since molecules and even atomic nuclei are nonrelativistic bound systems and the concept of potential energy is applicable to their components, the corresponding mass differences can be calculated using this concept. Thus, one can explain the released energy in terms of potential energy transforming into kinetic energy.

{\bf 10.8. Coulomb's law.} The binding energy of electrons in atoms is much lower than the electron mass. Hence, the concept of binding energy in atoms can be explained in terms of the nonrelativistic concept of potential energy. The binding energy $\varepsilon$ equals (with a minus sign) the sum of positive kinetic energy of the bound particle and its negative potential energy. The potential energy of, say, an electron in a hydrogen atom is given by Coulomb's law (in units, in which $\hbar$, $c=1$):
\be U= -{\alpha \over r}, \ee
where  $\alpha = e^{\,2}/\hbar c = 1/137$ and $e$ is the electron charge.

{\bf 10.9. More about potential energy.} The concept of potential energy is defined only in the Newtonian limit (see Landau and Lifshitz,{\it Mechanics}, [7]:  \S 5 ``The Lagrange function of a system of material points'' and \S 6 ``Energy''). The sum of kinetic and potential energies is conserved. If one of the two interacting particles is essentially relativistic, or both are, the concept of potential energy is inapplicable.

{\bf 10.10. Electromagnetic field.} The Coulomb field in the theory of relativity is the 0th component of the 4-potential of the electromagnetic field $A_i (i=0,1,2,3)$. The source of the field of a particle with electric charge $e$ is the 4-dimensional electromagnetic current
given in the next paragraph.
The interaction between two moving particles works through propagation of the field from one charge to the other. It is described by the so-called Green's function or the propagator of an electromagnetic field. (In quantum electrodynamics, we speak of propagation of virtual photons. The potential $A_i$ is a 4-vector because the spin of the photon equals unity.)

{\bf 10.11. Important clarification.} If a virtual photon carries away a 4-momentum $q$, then 4-momenta of the charged particle prior to the emission of a photon $p_{\rm in}$ and after its emission $p_{\rm fi}$ satisfy the condition $p_{\rm in} - p_{\rm fi} = q$. The 4-vector $p$ in the expression  $ep_i/E$ for the conserved current is $p=(p_{\rm in}+p_{\rm fi})/2$, and $E = \sqrt {E_{\rm in} E_{\rm fi}}$. As $p_{\rm in}^{\,2}=p_{\rm fi}^{\,2}=m^{\,2}$, so $qp=0$. (I denoted energy here by the letter $E$ because $e$ in the expression for current stands for charge. We are clearly short of letters.)

{\bf 10.12. Gluons and quarks.} A gluon's spin also equals unity. At first glance, the interaction between gluons and quarks is completely analogous to the interaction between photons and electrons. Not at second glance, though. The point is that all electrons carry the same electric charge while quarks have three different color charges. A quark emitting or absorbing a gluon may change its color. Clearly, this means that gluons must themselves be colored. It can be shown that there must be eight different color species of gluons. While photons are electrically neutral, gluons carry color charges.

{\bf 10.13. Quantum chromodynamics.} It might seem that color-charged gluons must be intense emitters of gluons, being a sort of `luminous light'. In fact, quantum chromodynamics --- the theory of interaction between quarks and gluons --- has a spectacular property known as confinement. In contrast to electrons and photons, colored quarks and gluons do not exist in a free state. These colored particles are locked `for life' inside colorless (white) hadrons. They can only change their incarceration locality. There are no Feynman diagrams with lines of free gluons or free quarks.

\section{Gravitation}

{\bf 11.1. Gravitational orbits.} Various emblems often show the orbits of electrons in atoms resembling the orbits of planets. It should be clear from the above that according to quantum mechanics, there are no such orbits in atoms. On the other hand, quantum effects are absolutely infinitesimal for macroscopic bodies, all the more so for such heavy ones as planets. Consequently, their orbits are excellently described by classical mechanics.

{\bf 11.2. Newton's constant.} The potential energy of the Earth in the gravitational field of the Sun is given by Newton's law
\be U = -{G M m\over r}, \ee
where $M$ is the solar mass, $m$ is the mass of the Earth, $r$ is distance between their centers, and $G$ is Newton's constant:
\be G = 6.71 \times  10^{-39} \hbar c \ \big[{\rm  GeV}/c^{\,2}\,\big]^{-2}.\ee
(Here we use units in which $c\ne 1$.)

{\bf 11.3. The quantity ${\bf p}_{\bf i}{\bf p}_{\bf k}/{\bf e}$.} The source of gravitation in Newtonian physics is mass. In the theory of relativity the source of gravitation is the quantity $p_ip_k/e$, which plays the role of a kind of `gravitational current'. (The reader will recall that $p_i$ is the energy-momentum 4-vector, and  $i = 0,1,2,3$. Consequently, the `gravitational current' has four independent components instead of the ten that a most general symmetrical four-dimensional tensor would have.)
The propagation of the field from the source to the `sink' is described by Green's function of the gravitational field or the propagator of the graviton --- a massless spin-2 particle. This propagator is proportional to $g^{\,il} g^{\,km} + g^{\,im} g^{\,kl} -g^{\,ik} g^{\,lm}$, where $g^{\,ik}$ is a metric tensor. (As in the case of the photon discussed above, the 4-momentum of the graviton is $q= p_{\rm in}-p_{\rm fi}$ and the 4-momentum in the expression for current is $p=(p_{\rm in} + p_{\rm fi})/2$, while $e=\sqrt {e_{\rm in} e_{\rm fi}}$. We are again short of letters! This time, letters for indices.)

{\bf 11.4. The graviton.} Like the photon, the graviton is a massless particle. This is the reason why Newton's and Coulomb's potentials have the form $1/r$. However, in contrast to the photon, which cannot emit photons, the graviton can and must emit gravitons. In this respect the graviton resembles gluons, which emit gluons.

{\bf 11.5. The Planck mass.} Elementary particle physics often uses the concept of the Planck mass:
\be m_{\rm P}=\sqrt{\hbar c\over G}.\ee


In units in which $c, \hbar =1$ we have $m_{\rm P}=1/\sqrt {G} = 1.22
\cdot10^{19}$ 
 GeV.

The gravitational interaction between two ultrarelativistic particles increases as the square of their energy $E$ in the center-of-inertia reference frame. It reaches maximum strength at $E \sim m_{\rm P}$ as the distance between the particles approaches $r\sim 1/ m_{\rm P}$. However, let us return from these fantastically large energies and short distances to apples and photons in gravitational fields of the Earth and the Sun.

{\bf 11.6. An apple and a photon.} Consider a particle in a static gravitational field, for instance, that of the Sun. The source of the field is the quantity $P_l P_m/ E$ where $P_l$ is the 4-momentum of the Sun and $E$ is its energy. In the rest frame of the Sun $l$, $m=0$ and $P_lP_m/E=M$, where $M$ is the solar mass. In this case the numerator of the propagator of the gravitational field $g^{\,il} g^{\,km} + g^{\,im} g^{\,kl} -g^{\,ik} g^{\,lm}$ is $2g^{\,i0}g^{\,k0} - g^{\,ik}g^{\,00}$, and the tensor quantity $p_i\,p_{ k}$ times the numerator of the propagator reduces to a simple expression $2e^{\,2} - m^{\,2}$. Hence, for a nonrelativistic apple of mass $m$ the `gravitational charge' equals $m$ while for a photon with energy $e$ it equals $2e$. Note the coefficient 2. Kinetic energy is attracted twice as strongly as the hidden energy locked in mass. This simple derivation of the coefficient 2 makes unnecessary the complicated derivation of paper [8] using isotropic coordinates.

{\bf 11.7. A photon in the field of the Sun.} The interaction of photons with the gravitational field must cause a deflection of a ray of light propagating from a remote star and passing close to the solar disk. In 1915 Einstein calculated the deflection angle and showed that it must be $4 GM/c^{\,2} R \simeq 1.75^{\,\prime\prime}$. (Here, $M$ and $R$ denote the solar mass and solar radius, respectively.) This prediction was confirmed during the solar eclipse of 1919, which stimulated a huge surge of interest in the theory of relativity.

{\bf 11.8. An atom in the field of the Earth.} As a nonrelativistic body on the Earth moves upwards, its potential energy increases in proportion to its mass. Correspondingly, the difference between energies of two levels of an atomic nucleus must be the higher, the higher the floor of the building in which this nucleus is located.

{\bf 11.9. A photon's energy is conserved.} On the other hand, the frequency $\omega$ of a photon propagating through a static gravitational field, and correspondingly its total energy $e=h\omega$, should remain unchanged.

As a result, a photon emitted on the ground floor of a building from a transition between two energy levels of a nucleus will be unable to produce a reverse transition in the same nucleus on the upper floor. This theoretical prediction was confirmed in the 1960s by Pound and Rebka [9] who used the just discovered M$\ddot{\rm o}$ssbauer effect, which makes it possible to measure the tiniest shifts in nuclear energy levels.

However, the wavelength changes. A photon propagating through a static gravitational field like a stone has its total energy $e$ and frequency $\omega$ conserved. However, its momentum and therefore wavelength change as the distance to the gravitating body changes.

{\bf 11.10. Refractive index.} As a photon moves away from the source of a gravitational field, its velocity increases and tends to $c$, and when it approaches the source, it decreases. Hence, the gravitational field, like a transparent medium, has a refractive index. This is a visually clear explanation of the deflection of light in the field of the Sun and in the gravitational lenses of galaxies. Shapiro experimentally discovered the decrease in the velocity of photons near the Sun when measuring the delay of the radar echo returned by planets.

{\bf 11.11. Clocks and gravitation.} Ordinary clocks, like atomic clocks, are ticking the faster, the higher they are lifted. Let two synchronized clocks A and B be placed on the first floor. If we move clock A to the second floor and then, say, a day later, move clock B to the second floor as well, clock A will be ahead of B as A has been ticking faster than B for 24 hours. Nevertheless, both A and B will continue to serve as identically reliable stopwatches.
When every point in space is assigned an individual clock, one in fact assumes that all clocks tick at a rate that is independent of the distance to gravitating bodies (in our case, on which floor of the building they are). However, this is not true for ordinary clocks. In order to distinguish extraordinary clocks from ordinary clocks, we will refer to extraordinary ones as `cloned'.
As we saw above, the frequency of light measured using clocks placed on various floors is independent of the floor number. If, however, it is measured with `cloned' local clocks, we discover that it is the lower, the higher is the floor. One interpretation of the Pound--Rebka experiment, stating that the energy of a vertically moving photon decreases with height, like the kinetic energy of a stone thrown upwards, is based on precisely this argument. However, a drop in kinetic energy of the stone is accompanied with an increase in its potential energy, so that the total energy is conserved. Now, a photon has no potential energy, so that its energy in a static gravitational field remains constant.

\section{Epistemology and linguistics}

{\bf 12.1. Physics and epistemology.} {\it Episteme} in Greek means knowledge. Epistemics is the science of knowledge, a relatively young branch of epistemology, the theory of knowledge and cognition. Obviously, the problems I discuss in this talk concern not only physics but epistemology, too.

{\bf 12.2. Physics and semantics.} The Greek attribute `semanticos' (signifying) was used in linguistics already by Aristotle. However, what are the links tying the science of languages --- linguistics --- and semantics --- the science of words and symbols, an element of linguistics --- to physics?
This is the right moment to recall the words allegedly said by V A Fock: ``Physics is an essentially simple science. The most important problem in it is to understand what each letter denotes.''
XX century physics drastically changed our understanding of what a vacuum and matter are, and connected in a new way such properties of matter as energy, momentum, and mass. The elaboration of the fundamental concepts of physics has not been completed and is unlikely to end in the foreseeable future. This is one of the reasons why it is so important to choose the adequate words and letters when discussing physical phenomena and theories.

{\bf  12.3. Concepts glued together'.} Newton's {\it Principia} `glued together' the concepts of mass and matter (substance): ``mass is proportional to density and volume.'' In Einstein's papers mass is `glued together' with inertia and gravitation (the inertial and gravitational masses). And energy is glued to matter.

{\bf 12.4. The archetype.} According to dictionaries, an archetype is the historically original form (the protoform), the original concept or word, or the original type (prototype). The concept of the archetype keenly interested Pauli, who in 1952 published a paper on the effect of archetypical notions on the creation of natural-science theories by Kepler. It is possible that the concept of mass is just the archetypical notion that glued together the concepts of matter, inertia, and weight.

{\bf 12.5. Atom and archetype.} {\it Atom and Archetype} --- that was the title chosen for the English translation from German of the book [10] presenting the correspondence between Wolfgang Pauli and the leading German proponent of psychoanalysis Carl Jung, covering the period from 1932 to 1958. W. Pauli and C. Jung discussed, among other things, the material nature of time and the possibility of communicating with people who lived several centuries or millennia before us. It is widely known that Pauli treated rather seriously the effect named after him: when he walked into an experimental laboratory, measuring equipment broke down.

{\bf 12.6. Poets on terminology.} David Samoilov on words: ``We wipe them clean as we clean glass. This is our trade.'' Vladimir Mayakovskii: ``The street is writhing for want of tongue. It has no nothing for yelling or talking.'' ({\it Translated by Nina Iskandaryan.})
Many an author responds to the dearth of precise terms and inability to use them by resorting to meaningless words like `rest mass' which impart smoothness and `energetics' to texts, just as `blin' \footnote{`Blin' is a slang euphemism for a `four-letter word' in vulgar Russian.} does to ordinary speech.

{\bf 12.7. How to teach physics.} Terms need `wiping clean' and `unglueing'.
The `umbilical cord' connecting the modern physical theory with the preceding `mother theory' needs careful cutting in teaching. (In the case of the theory of relativity the mother was the `centaur' composed of Maxwell's field theory and Newton's mechanics, with relativistic mass serving as the umbilical cord.)
Let us recall the title of F Klein's famous book {\it Elementary Mathematics from an Advanced Standpoint}. The landscape of modern physics must be contemplated from an advanced standpoint: not from a historical gully but from the pinnacle of symmetry principles. I firmly believe that it is unacceptable to claim that the dependence of mass on velocity is an experimental fact and thus hide from the student that it is a mere interpretational `factoid'. (Dictionaries explain that a factoid looks very much like a fact but is trusted only because we find it in printed texts.)

\section{Concluding remarks}

{\bf 13.1. The ${\bf `E}={\bf mc}^{\,\bf 2}$ problem': could it be avoided?} One is tempted to think that the `$E=mc^{\,2}$ problem' would not arise from the first place if the quantity $E/c^{\,2}$ --- the proportionality coefficient between velocity and momentum --- were identified with a new physical quantity christened as, say, `inertia' or `iner'; it would be identical to mass as momentum tended to zero. As a result, mass would become `rest inertia'. Likewise, another new quantity could be introduced --- `heaviness' or `grav' --- $p_ip_k/E$ reducing to mass at zero momentum. But physicists preferred `to refrain from multiplying entities' and from introducing new physical quantities. They formulated instead new, more general relations between old quantities, for example $E^{\,2}-{\bf p}^{\,2} c^{\,2}=m^{\,2}c^{\,4}$ and  ${\rm \bf p} = {\rm \bf v}E/c^2$.

Unfortunately, many authors attempt to retain even in relativistic physics such nonrelativistic equations as ${\bf p}=m{\bf v}$, and such nonrelativistic glued-up concepts as `mass is a measure of inertia' and `mass is a measure of gravitation'; as a result, they prefer to use the notion of velocity-dependent mass.
It is amazing how again and again a physicist would choose the first of these paths (new equations) in his research papers and the second one (old glued-up concepts) in science-popularizing and pedagogical activities. This could of course only produce unbelievable confusion in the minds of those who read popular texts and blindly follow the authority.

{\bf 13.2. On the reliability of science.} An opinion that has become widely publicized recently is that science in general and physics in particular are untrustworthy. Many popularizers of science create the impression that the theory of relativity proved Newton's mechanics wrong just as chemistry proved alchemy wrong and astronomy proved astrology wrong. Such declarations are a crude distortion of the essence of scientific revolutions. Newton's mechanics remains a correct science today, in the XXI century, and will continue to be correct forever. The discovery of the theory of relativity only put bounds on the domain of applicability of Newton's mechanics to velocities much smaller than the speed of light $c$. It also demonstrated its approximate nature in this domain (to within corrections of the order of $v^{\,2}/c^{\,2})$.
Similarly, the discovery of quantum mechanics put bounds on the domain of applicability of classical mechanics to phenomena for which the quantity of action is large in comparison with the quantum of action $\hbar$. Quite to the contrary, the domain where astrology and alchemy exist is that of prejudice, superstition, and ignorance. It is rather funny that those who compare Newton's mechanics with astrology typically believe that mass depends on velocity.

{\bf 13.3. Recent publications.} Additional information on the aspects discussed above can be found in [11, 12].

{\bf 13.4. On the title.} My good friend and expert in the theory of relativity read the slides of this talk and advised me to drop Pythagoras's name from the title. I chose not to follow his advice as in the relativity-related literature I had never come across a discussion of right-angled triangles without the approximate extraction of square roots.

{\bf 13.5. Acknowledgment} I am grateful to T Basaglia, A Bettini, S I Blnnikov, V F Chub, M A Gottlieb E G Gulyaeva, E A Ilyina, C Jarllscog, V I Kisin, B A Klumov, B L Okun,  S G Tikhodeev, M B Voloshin, V R Zoller for their advice and help. The work was supported by the grants NSh-5603.2006.2, NSh-4568.2006.2 and RFBR-07-02-00830-a.


\bigskip

\centerline{\bf References}

\bigskip

[1]	Okun L B ``Formula Einshteina: $E_0=mc^{\,2}$. `Ne smeetsya li Gospod' Bog'?'' {\it Usp. Fiz. Nauk} {\bf 178} 541 (2008) [``The Einstein formula $E_0=mc^{\,2}$ `Isn't the Lord laughing'?'' {\it Phys. Usp.} {\bf 51} 513 (2008)], arXiv:0808.0437

[2]	Klein F Elementarmathematik vom h$\ddot{\it o}$eheren Standpunkte aus. Erster Band, Arithmetik, Algebra, Analysis 3 Auflage (Berlin: Verlag fon Julius Springer, 1924) [Translated into English: Elementary Mathematics from an Advanced Standpoint, Arithmetics, Algebra, Analysis (New York: Dover Publ., 2007); Translated into Russian (Moscow: Nauka, 1987)]

[3]	Poincar\'{e} H "Sur la dynamique de l'electron", {\it Rendiconti del Circolo Matematico di Palermo} {\bf 21} 129 (1906) [Translated into Russian: ``O dinamike elektrona'' (``On the dynamics of electron'') Izbrannye Trudy (Selected Works) Vol. 3 (Moscow: Nauka, 1974) p. 433]

[4]	Minkowski H "Raum und Zeit" {\it Phys. Z.}, {\bf 10}, 104--111 (1909) \rm [Translated into Russian: ``Prostranstvo i Vremya'' (``Space and time''), in Lorentz H A, Poincare H, Einstein A, Minkowski H {\it Printsip Otnositel'nosti. Sbornik Rabot Klassikov Relyativizma} (The Principle of Relativity. Collected Papers of Classics of Relativism) (Eds V K Frederiks, D D Ivanenko) (Moscow -- Leningrad: ONTI, 1935) pp. 181 -- 203]

[5]	Hawking S {\it The Universe in a Nutshell} (New York: Bantam Books, 2001) [Translated into Russian (Translated from English by A Sergeev) St.-Petersburg: Amfore, 2007)]

[6]	Landau L D, Lifshitz E M {\it Teoriya polya} (The Classical Theory of Fields) (Moscow: Nauka, 1988) [Translated into English (Amsterdam: Reed Elsevier, 2000)]

[7]	Landau L D, Lifshitz E M {\it Mekhanika} (Mechanics) (Moscow: Nauka, 1988) [Translated into English (Amsterdam: Elsevier Sci., 2003)]

[8]	Okun L B  ``The concept of mass'' {\it Phys. Today} {\bf 42} (6) 31--36 (1989); Okun L B ``Putting to rest mass misconceptions'' {\it Phys. Today}  {\bf 43} (5) 13, 15, 115, 117 (1990)

[9]	Pound R V, Rebka G A (Jr.) ``Apparent weight of photons'' {\it Phys. Rev. Lett.} {\bf 4} 337--341 (1960)

[10]	Pauli W, Jung C {\it Atom and Archetype: Pauli/Jung Letters. 1932--1958.} (Ed. C A Meyer) (Princeton, NJ: Princeton Univ. Press, 2001); Meier C A (Herausgegeben)  Wolfgang Pauli, C. G. Jung {\it Ein Briefwechsel 1932--1958}  Berlin: Springer-Verlag, (1992)

[11]	Okun L B ``Chto takoe massa? (Iz istorii teorii otnositel'nosti)'' (`What is mass? (From the history of relativity theory)'), in {\it Issledovaniya po Istorii Fiziki i Mekhaniki.} 2007 (Research on the History of Physics and Mechanics. 2007) (Executive Ed. G M Idlis) (Moscow: Nauka, 2007)

[12]	Okun L B ``The evolution of the concepts of energy, momentum and mass from Newton and Lomonosov to Einstein and Feynman'', in {\it Proc. of the 13th Lomonosov Conf. August 23, 2007} (Singapore: World Scientific) (in press)


\end{document}